# Assessing Cardiac Dynamics through RF Sensing for Hemodynamic Monitoring in Pacemakers

A. Khaleghi[1], J. Bergsland[2], I. Balasingham[12]

[1]Department of Electronic Systems, Norwegian University of Science and Technology, Norway
[2]Intervention Center, Oslo University Hospital, Norway
ali.khaleghi@ntnu.no

***Abstract* — This paper examines the use of radiofrequency (RF) channels for hemodynamic monitoring in cardiac pacemakers. It analyzes RF signal variations between intracardiac transceivers in the right ventricle (RV) and right atrium (RA), as well as subcutaneous receivers, to determine their correlation with cardiac dynamics. The study shows that temporal RF signal variations closely align with cardiac rhythm, allowing for the estimation of parameters such as chamber volume, valve behavior, and pressure changes. These results underscore the potential of RF-based sensing as a novel method for real-time cardiac monitoring in pacemaker systems.**

## I. Introduction

Cardiac pacemakers are indispensable in managing heart failure (HF) and other cardiovascular conditions, significantly improving survival and quality of life [1], [2]. . Widely used to treat heart block, arrhythmias, and HF, pacemakers are available in configurations ranging from single-chamber systems to advanced biventricular devices for cardiac resynchronization therapy (CRT) (Fig.1). These systems enhance cardiac efficiency and synchrony, addressing the complexities of HF. Modern pacemakers also integrate sensors such as accelerometers and impedance monitors, enabling adaptive responses to physiological demands. However, their primary focus remains rhythm regulation, with limited capabilities for real-time monitoring of critical hemodynamic parameters.

Hemodynamic monitoring is essential in HF management, providing real-time insights into cardiac function that enable timely interventions and improve patient outcomes. Integrating hemodynamic monitoring into pacemaker systems could further enhance HF care by offering continuous, real-time data on parameters such as pulmonary artery pressure (PAP) and left atrial pressure (LAP). This integration would allow for proactive therapeutic adjustments, potentially reducing hospitalizations and improving quality of life for HF patients. Advancements in implantable hemodynamic monitors, such as the CardioMEMS HF System and the Vectorious LAP Monitor, have demonstrated the value of continuous monitoring in HF management [3], [4], [5]. Incorporating similar capabilities into pacemaker systems, particularly through non-invasive enhancements that seamlessly integrate with existing implanted devices, presents a promising approach for comprehensive cardiac care.

This paper proposes the integration of radiofrequency (RF) technology into pacemakers to enable hemodynamic monitoring. By analyzing RF channel characteristics, the proposed approach assesses cardiac parameters such as chamber volume, motion, blood perfusion, and pressure. Utilizing non-ionizing RF frequencies, this method enhances pacemaker functionality with minimal modifications to existing systems, advancing HF management.

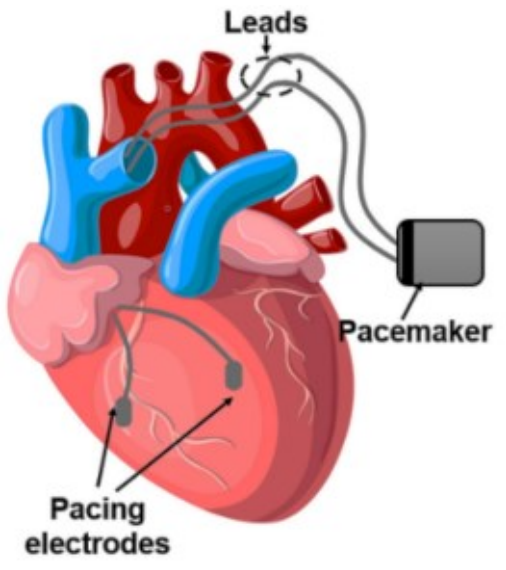

Fig.1 Dual chamber lead based cardiac pacemaker with subcutaneous electronic pacemaker can

## II. Radio Frequency Measurement Setup

The measurement system utilized in this study relies on RF signals transmitted through bipolar cardiac pacemaker leads, which are typically inserted through a vein into the heart during the pacemaker implantation procedure (Fig.2). Lead placement depends on the type of pacemaker and the patient's specific needs. For a single-chamber pacemaker, one lead is placed in either the right atrium or ventricle. Dual-chamber pacemakers use two leads positioned in the right atrium and ventricle, while biventricular pacemakers may include a third lead to synchronize the pumping action of both ventricles. These leads are carefully positioned within the heart to ensure reliable delivery of electrical impulses to the appropriate chambers. Pacemaker leads not only deliver electrical signals to stimulate heart contractions and regulate the heartbeat, but they also transmit information about the heart's electrical activity back to the pacemaker device, allowing it to monitor and adjust pacing as needed.

For our research, we utilized commercially available bipolar pacemaker leads (Medtronic CapSureFix Novus MRI™ SureScan® 5076) surgically implanted in the right atrium (RA) and right ventricle (RV) of a pig's heart. The 5076 lead, designed for both atrial and ventricular applications, features a retractable, electrically active helix tip with a 1.17 mm diameter and a 1.8 mm length for secure myocardial attachment and precise positioning. Its 2.0 mm lead body is insulated with treated silicone rubber for durability and flexibility. The conductor, made from MP35N alloy (a nickel-cobalt-chromium-molybdenum material), provides reliable electrical conductivity essential for stable pacing and sensing. The lead is equipped with an IS-1 BI connector, ensuring compatibility with various implantable pulse generators. Figure 2 illustrates the intricate structure of the lead, including the retractable helix screw mechanism and compact design. The bipolar cable consists of an active pole (spring tip) and a ground connection approximately 1 cm from the tip screw, with the outer shield connected to the ground while the cable remains insulated from tissue contact.

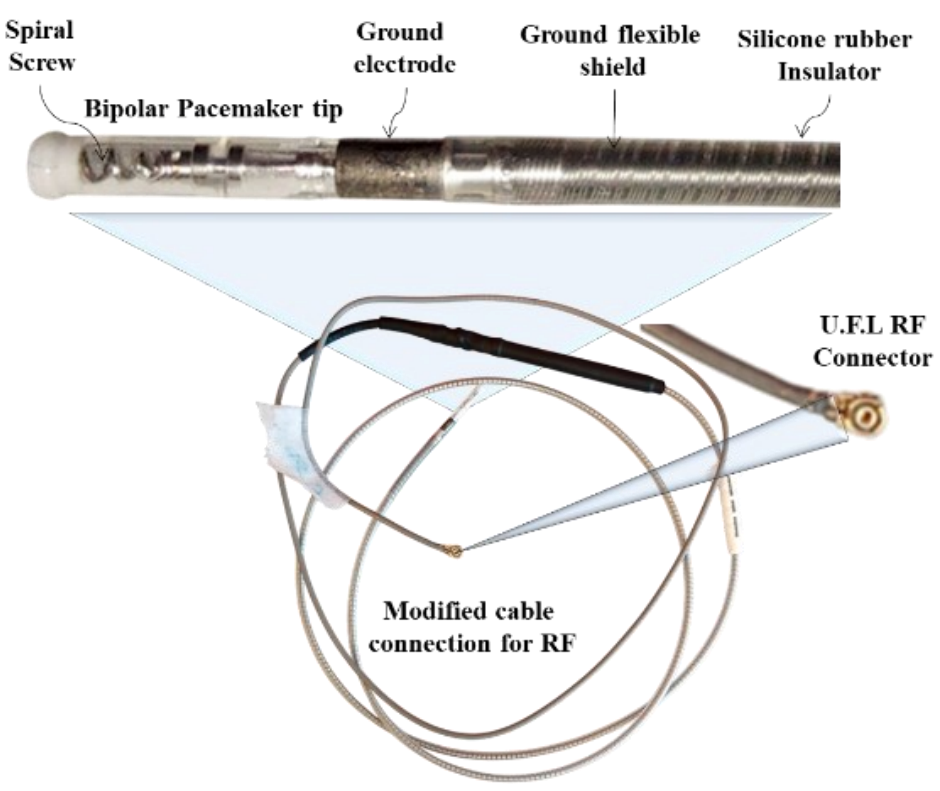

Fig. 2 Medtronic bipolar pacemaker leads employed for low-frequency measurements of transmitters and receivers within the cardiac chambers of the pig model. (a) a general view of the bipolar pacemaker cable, and b) modified connector to link to RF measurement equipment

The pacemaker leads were modified for RF transmission by adapting the input connector to match a U.FL RF cable, allowing signal transmission from an RF signal generator and reception by a spectrum analyzer (Fig. 2) during experiments. In an integrated design, the RF signal can be routed without interference with the electrical pacing signals on the same cable, and the RF connector can be matched to the existing standard pacemaker can connector. Measurements using the Medtronic cable were conducted at 200 kHz and 1 MHz to evaluate RF channel variations within the low-frequency spectrum. Frequencies above 1 MHz led to significant signal leakage and poor reception at the receiver, making lower frequencies more reliable and practical. Furthermore, operating at lower frequencies offers improved energy efficiency, making it particularly advantageous for battery-operated signal generators.

The RF signal was generated at 200 kHz and 1 MHz with a power level of +10 dBm on a 50-ohm impedance and transmitted through leads implanted in the RA and RV. The impedance of the leads at these frequencies was measured to be approximately 100 ohms (real values).

The signal was received subcutaneously at two chest locations, simulating potential placements for the pacemaker can. The subcutaneous receiver setup consisted of a bipolar electrode with two patches, each equipped with 4 mm × 4 mm metal contacts, spaced 1 cm apart and matched to a U.FL RF cable connected to a spectrum analyzer (R&S FSH4) as the signal receiver.

Signal fluctuations were recorded in zero-span mode, with the receiver frequency set to match the transmitter frequency. The bandwidth was set to 100 Hz, with a scan time of 5 seconds per measurement, recorded over a total of 30 seconds to ensure sufficient data for signal analysis.

The receiver impedance is 50 ohms, while the implant electrode impedance is approximately 100 ohms at the test frequencies. The impedance mismatch affects only the accepted signal level during transmission and the received signal levels at the receiver but does not impact signal variations.

The open-chest surgery was conducted on a 65 kg pig subject, to place the pacemaker cables, after which the incision was fully closed to stabilize cardiac motion. The procedure was performed under general anesthesia at the intervention center of Oslo University Hospital. Throughout the experiment, the heart maintained its rhythmic beating, with dynamic cardiac activity visibly influencing the recorded RF signals (Fig.3). Additional sensors for ECG, as well as aortic, atrial, and ventricular blood pressure, were implanted to facilitate comprehensive data acquisition. The NI PowerLab system was used for data acquisition from the cardiac sensors, allowing measurement of physiological signals. The timing of the PowerLab system and spectrum analyzer was synchronized using a common reference clock. Aortic pressure (CH1), RA pressure (CH2), RV pressure (CH3), and ECG (CH4) were concurrently captured and synchronized. The recorded data were subsequently used for post-processing and analysis. This approach allowed evaluation of correlations between RF signals and cardiac functions.

## III. Results

### A. RA to subcutaneous receiver observations

The normalized received signal variations over time at 200 kHz are shown in Fig. 4. The average received signal level is -43.9 dBm, with a variation range of 2.33 dB. Changing the receiver electrode placement to different subcutaneous locations affects only the average signal level, while the variation range remains consistent. Measurements at 1 MHz exhibit a stronger average signal level of -40.9 dBm and a reduced variation range of 1.47 dB. The measured cardiac hemodynamics, including ventricular pressure, atrial pressure, and EEG, are plotted synchronously with the RF signal. Synchronization accuracy is within ±50 milliseconds, limited by the setup configuration and the integration of two different measurement systems: PowerLab and the spectrum analyzer. The measured curves are normalized to unity and shifted along the Y-axis for comparative cardiac hemodynamic analysis.

The observed RF signal variations correlate precisely with the dynamics of RA blood volume, RA pressure, and valve status throughout the cardiac cycle. The RF signal is at its minimum during the T-wave, reflecting maximum blood volume in the RA as it fills with blood returning from the systemic veins. During this phase, RA pressure is low but rising gradually, and the tricuspid valve remains closed due to higher RV pressure. Following this, the RF signal increases sharply to a first peak between the end of the T-wave and the start of the P-wave, marking the passive emptying of the RA into the RV during early diastole. The RA blood volume decreases as blood flows rapidly through the open tricuspid valve, while RA pressure rises slightly as ventricular filling progresses. The RF signal then dips slightly during mid-diastole, corresponding to the diastasis phase, where blood flow between the RA and RV slows as pressures in both chambers equalize. During this phase, RA blood volume stabilizes, and RA pressure remains low and steady. As the P-wave begins, the RF signal increases again to a second, higher peak, driven by atrial contraction (atrial kick). During this phase, the RA actively contracts, pushing additional blood into the RV. This causes a sharp decrease in RA blood volume, while RA pressure peaks, reflecting the active force of contraction.

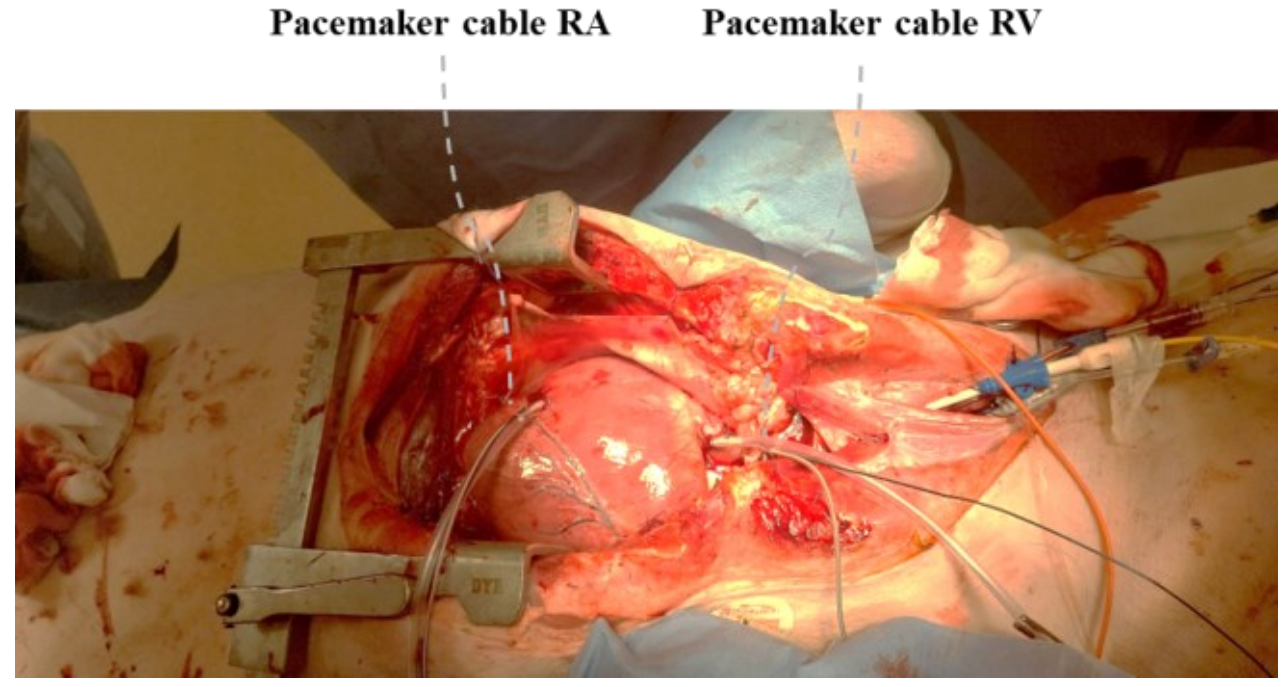

Fig.3 Open chest surgery for placement of the pacemaker cables in the RV and RA

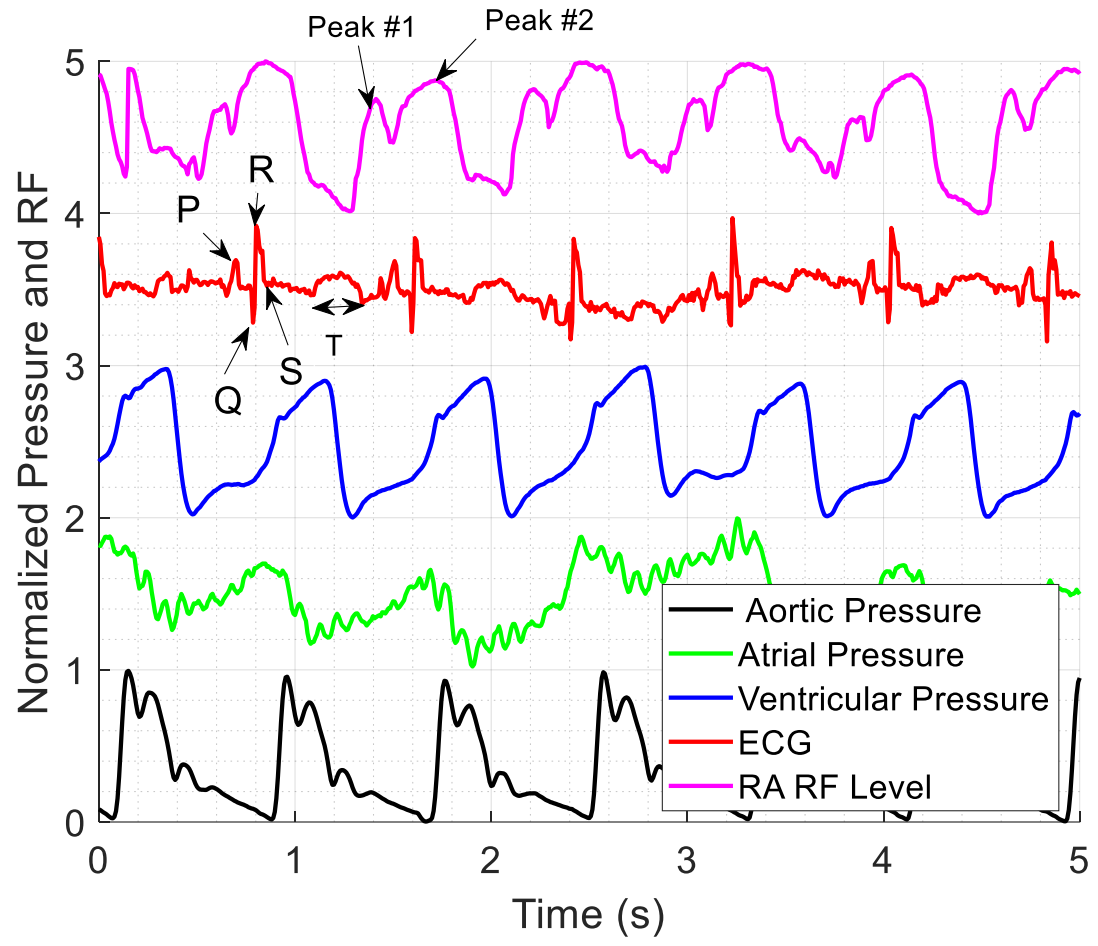

Fig. 4 Measured RF signal transmitted from the RA to a subcutaneous receiver electrode, recorded alongside EEG, ventricular pressure, and atrial pressure. Measured using RA Medtronic pacemaker transmitter and received with subcutaneous electrode at 200 kHz.

After the second peak, the RF signal flattens, indicating that the RA has emptied most of its blood and remains relatively low in volume. This stability continues until ventricular systole begins. Following the QRS complex, the RF signal decreases sharply in two steps during ventricular systole. The first drop occurs during isovolumetric contraction, where retrograde pressure from the RV causes RA pressure to rise, while the tricuspid valve closes to prevent backflow. The second drop occurs as the RA begins to fill rapidly during late systole, with blood returning from the systemic veins. By the end of the T-wave, the RA reaches its maximum blood volume, and RA pressure begins to rise again. These observations align well with the cardiac cycle and the interplay of atrial and ventricular dynamics.

### *B. RV to subcutaneous receiver observations*

The RF signal, was applied to the RV bipolar pacemaker electrode. The received signal was recorded via a subcutaneous electrode placed at two chest locations (Fig.5). At 200 kHz, the average received signal level was -34 dBm, with a variation range of 1.14 dB, while at 1 MHz, the signal level increased to -30 dBm with a variation range of 1.02 dB. The measured results are presented in Figure 5.

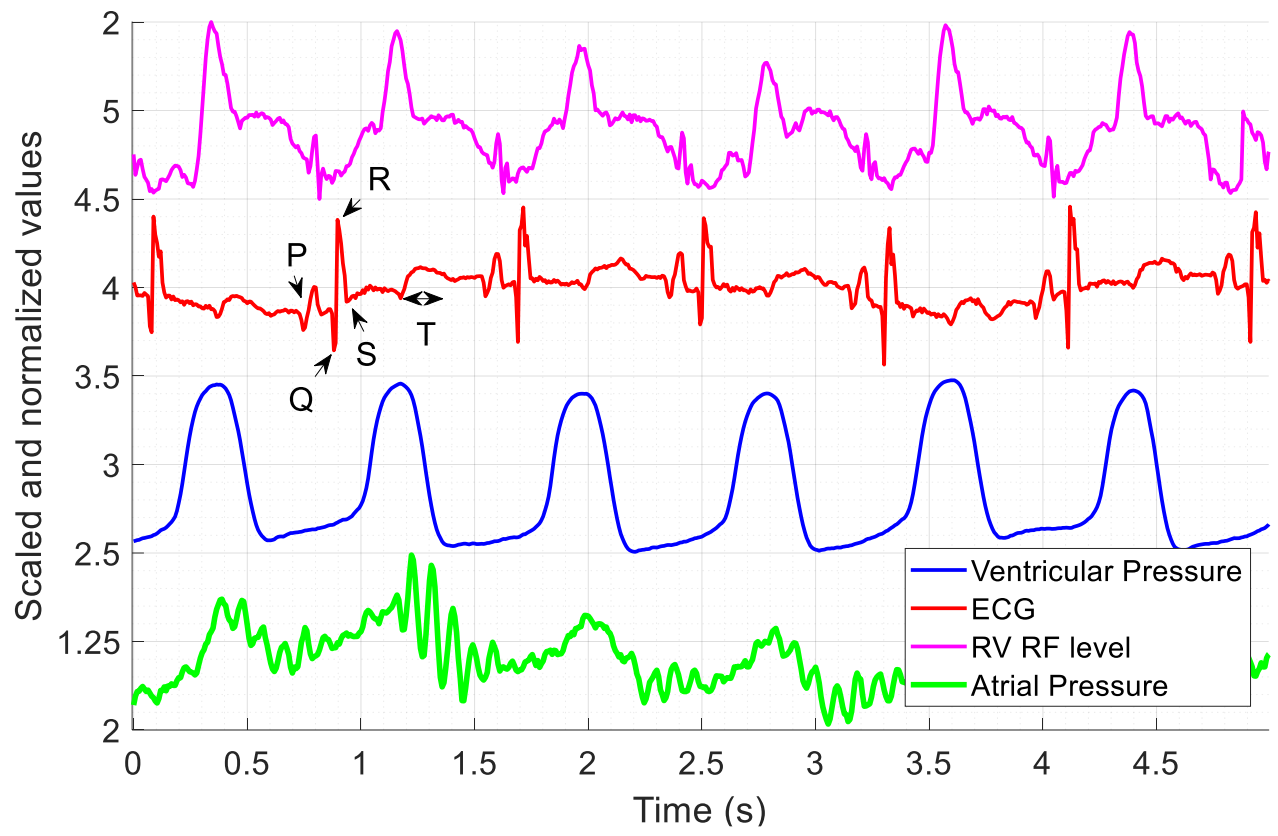

Fig. 5 Measured RF signal transmitted from the RV to a subcutaneous receiver electrode, recorded alongside EEG, ventricular pressure, and atrial pressure. Measured using RV Medtronic pacemaker electrode with signaling at 200 kHz and received with subcutaneous electrode.

The observed RF signal variations from the RV, correlated with the cardiac cycle, provide detailed insights into the interplay between RV pressure, blood volume, and valve status. Just after the QRS complex, the RF signal is at its minimum, indicating maximum RF signal loss, which corresponds to maximum RV blood volume (end-diastolic volume, EDV) as the ventricle is fully filled. During this phase, RV pressure begins to increase rapidly as the ventricle enters isovolumetric contraction, with both the tricuspid and pulmonary valves closed. As the heart transitions to mid-ST, the RF signal loss reduces sharply, reflecting rapid blood ejection from the RV. This occurs as the pulmonary valve opens, allowing blood flow into the pulmonary artery, and RV pressure reaches its peak during systolic contraction. From mid-ST to the start of the T-wave, the RF signal loss reduces further, reaching its minimum as the RV blood volume decreases to its lowest point (end-systolic volume, ESV), while RV pressure starts to decline.

During the T-wave, the RF signal loss increases rapidly, corresponding to early diastolic filling as the tricuspid valve opens. The RV blood volume increases sharply, driven by a significant pressure gradient between the RA and RV, while RV pressure continues to decrease rapidly during this isovolumetric relaxation phase. By mid-T-wave, the RF signal loss stabilizes, aligning with the diastasis phase, where blood flow into the RV slows as the RA and RV pressures equalize. Following the P-wave, the RF signal loss increases further due to atrial contraction (atrial kick), pushing additional blood into the RV, and RV pressure increases slightly as the RV reaches its maximum blood volume (EDV).

Additionally, a short sine wave fluctuation in RF loss is observed during the P-wave and just before the QRS complex. This phenomenon likely reflects transient oscillations in RV pressure and dynamic blood flow adjustments caused by atrial contraction. These fluctuations are consistent with momentary flow variations across the tricuspid valve, which remains open and dynamically adjusts during atrial systole. These observations collectively confirm the correlation between RF signal dynamics, RV blood volume, RV pressure, and valve behavior throughout the cardiac cycle.

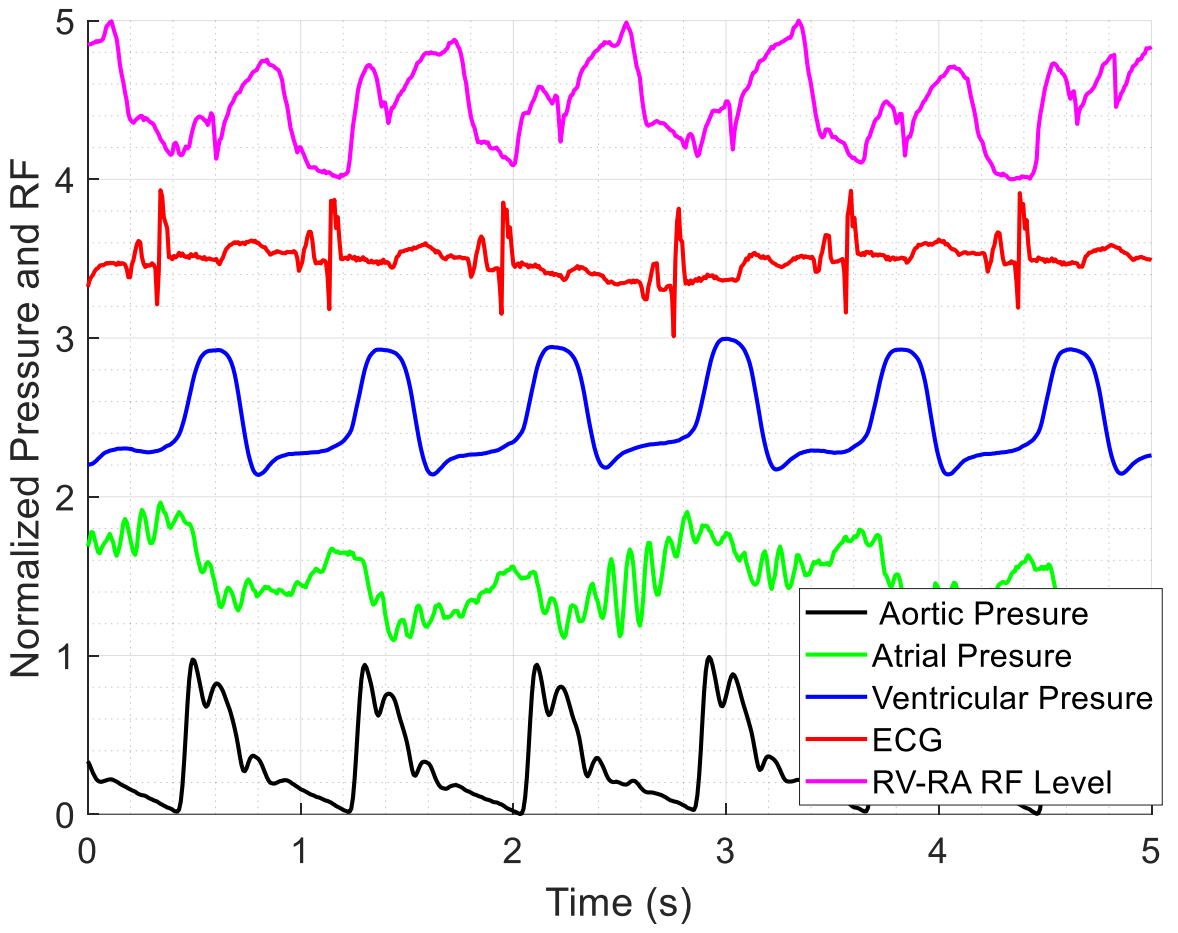

Fig. 6 Measured RF signal transmitted from the RA pacemaker electrode and received by the RV electrode, recorded alongside EEG, ventricular pressure, atrial pressure and Aortic pressure, frequency at 200 kHz.

### C. *RV to RA channel observations*

The RF signal, was applied to the RA bipolar pacemaker electrode. The received signal was recorded via the second pacemaker electrode at RV for dual chamber channel pacemaker scenario. At 200 kHz, the average received signal level was -47.8 dB with a variation range of 2.4 dB, while at 1 MHz, the signal level increased to -43.7 dBm with a variation range of 1.53 dB.

The observed RF signal variations between the RV and RA (Fig.6) closely correlate with the dynamics of their combined blood volume and the cardiac cycle. At the S-point of the ECG, the RF signal is at its minimum, reflecting the maximum combined blood volume in the RV and RA. This corresponds to isovolumetric contraction, where the RV is at its end-diastolic volume (EDV), and the RA is maximally filled. The signal then increases rapidly to a first peak during the mid-ST segment, as the RV begins rapid ejection, reducing its volume, while the RA volume increases slightly due to venous return. By the start of the T-wave, the RF signal decreases, corresponding to reduced ejection in the RV and steady filling of the RA, causing a temporary stabilization or slight increase in combined blood volume.

During the T-wave, the RF signal increases again, reaching a higher level by the end of the T-wave, as the RV approaches its end-systolic volume (ESV) and the RA is nearly maximally filled. Following this, the RF signal increases slightly and stabilizes just before the P-wave, reflecting a balance during early diastole, where the RV fills rapidly, and the RA empties partially. As the P-wave begins, the RF signal decreases, driven by atrial contraction, which actively transfers blood from the RA to the RV, increasing combined blood volume. This decrease continues into the QRS complex, where the RV reaches its maximum volume (EDV), completing the cycle. These variations align with the expected blood volume dynamics and valve operations throughout the cardiac cycle.

## IV. Conclusion

This study demonstrates the feasibility and potential of integrating radiofrequency (RF) technology into pacemaker systems for hemodynamic monitoring. By analyzing RF signal variations in three configurations—RV to subcutaneous, RA to subcutaneous, and RV to RA links—we established strong correlations between RF channel dynamics and cardiac hemodynamics. The RF signal variations were shown to reflect critical cardiac parameters such as chamber volume, blood perfusion, and valve behavior, aligning closely with measured ventricular and atrial pressures as well as ECG data.

The results highlight that RF-based sensing can be seamlessly incorporated into existing pacemaker systems, offering non-invasive, real-time monitoring of cardiac function. This integration could enable proactive therapeutic adjustments, improve patient outcomes.

## Acknowledgment

The project got funding from research Council Norway of Norway, IKTpluss program, Project WINNOW and EU Project of B-CRATOS FET OPEN Grant No. 96504.